\documentclass[conference]{IEEEtran}
\IEEEoverridecommandlockouts
\usepackage{cite}
\usepackage{amsmath,amssymb,amsfonts}
\usepackage{algorithmic}
\usepackage{graphicx}
\usepackage{textcomp}
\usepackage{xcolor}
\usepackage{ulem}
\usepackage{soul}
\usepackage{hyperref}
\usepackage{subcaption}

\def\BibTeX{{\rm B\kern-.05em{\sc i\kern-.025em b}\kern-.08em
    T\kern-.1667em\lower.7ex\hbox{E}\kern-.125emX}}

\makeatletter
\newcommand{\linebreakand}{%
  \end{@IEEEauthorhalign}
  \hfill\mbox{}\par
  \mbox{}\hfill\begin{@IEEEauthorhalign}
}
\makeatother

\begin{document}

\title{Likelihood ratio map for direct exoplanet detection
\thanks{This work was supported by the Fonds de la Recherche Scientifique - FNRS and the Fonds Wetenschappelijk Onderzoek - Vlaanderen under EOS Project no. 30468160. Simon Vary is a beneficiary of the FSR Incoming Post-doctoral Fellowship.}
}
\author{\IEEEauthorblockN{Hazan Daglayan} 
\IEEEauthorblockA{\textit{ICTEAM Institute} \\ 
\textit{UCLouvain}\\
 Louvain-la-Neuve, Belgium\\ 
hazan.daglayan@uclouvain.be} 
\and
\IEEEauthorblockN{Simon Vary} 
\IEEEauthorblockA{\textit{ICTEAM Institute} \\ 
\textit{UCLouvain}\\ 
 Louvain-la-Neuve, Belgium\\ 
simon.vary@uclouvain.be} 
\and
\IEEEauthorblockN{Faustine Cantalloube} 
\IEEEauthorblockA{\textit{CNRS, CNES, LAM} \\ 
\textit{Aix Marseille Univ}\\ 
Marseille, France\\  
faustine.cantalloube@lam.fr} 
\and 
\linebreakand
\IEEEauthorblockN{P.-A. Absil} 
\IEEEauthorblockA{\textit{ICTEAM Institute} \\ 
\textit{UCLouvain}\\ 
 Louvain-la-Neuve, Belgium\\ 
pa.absil@uclouvain.be} 
\and
\IEEEauthorblockN{Olivier Absil} 
\IEEEauthorblockA{\textit{STAR Institute} \\ 
\textit{Universit\'e de Li\`ege}\\ 
Li\`ege, Belgium \\ 
olivier.absil@uliege.be} 

}

\maketitle

\begin{abstract}
Direct imaging of exoplanets is a challenging task due to the small angular distance and high contrast relative to their host star, and the presence of quasi-static
noise. We propose a new statistical method for direct imaging of exoplanets based on a likelihood ratio detection map, which assumes that the noise after the background subtraction step obeys a Laplacian distribution. We compare the method with two detection  approaches based on signal-to-noise ratio (SNR) map after performing the background subtraction by the widely used Annular Principal Component Analysis (AnnPCA). The experimental results on the Beta Pictoris data set show the method outperforms SNR maps in terms of achieving the highest true positive rate (TPR) at zero false positive rate (FPR).
\end{abstract}


\section{Introduction}
Out of the nearly 5000 exoplanets that have been discovered, most of them in recent years, only around a tenth have been detected by methods of \textit{direct imaging}, that is, based on the faint light they emit \cite{nasa}. In fact, the existence of most of the exoplanets is verified by \textit{indirect imaging} methods based on measuring the effect a planet has on the starlight reaching Earth, such as when the planet blocks the light of the star, called the \textit{transit method}. However, indirect methods are limited to a specific alignment between the planet, the star, and the observer, and biased towards close and massive planets orbiting old quiet stars.

Directly imaging an exoplanet is a challenging task due to the light emitted by the planet being very faint and its resolution being very small, especially compared to the light emitted by the nearby star, referred to  as the \textit{host}. This motivates the need for telescopes capable of both high resolution and high contrast, which due to physical constraints, are limited to being ground-based rather than space-based. Consequently, the images obtained by the ground-based telescopes are deformed by the atmospheric turbulence and other instrumental aberrations of the optics resulting in high-intensity noise called \textit{quasi-static speckles}, whose shape and intensity are inconveniently similar to the planet companions we are trying to find. 

\textit{Angular differential imaging} (ADI) partially overcomes the problem of quasi-static speckles by taking a sequence of images of a target star over a single night of observation without compensating for the Earth's rotation \cite{marois2006}. As a result, potential planet companions end up following circular trajectories across the image sequence, while the star and its quasi-static speckle field, being almost fixed with respect to the pupil of the telescope, remain roughly in the same location, exposing the detection problem to the use of dynamic-foreground/static-background separation methods \cite{bouwmans2017}.

The complete pipeline of ADI detection consists of three steps. The first is the background subtraction, which estimates the \textit{model PSF} containing the speckles and the \textit{residual cube} that is meant to contain only the planets and some residual noise. The standard and most widely used methods are based on low-rank matrix models, such as principal component analysis (PCA) \cite{amaraquanz2012,soummer2012}, its annular version (AnnPCA) \cite{gonzalezvip2017}, and the low-rank plus sparse method (LLSG)~\cite{gonzalezllsg2016}. The rationale behind the low-rank models is that the bright, quasi-static speckles are captured by the first few principal components, while the higher-rank moving planets are excluded from the model. Other methods are based on a maximum likelihood approach \cite{cantalloube2015} or supervised machine learning methods \cite{gonzalezsupervised2018}. 

The second step is the flux estimation of the residual cube, in which  we, under some probabilistic model, compute the \textit{flux}, i.e.~the estimated light intensity of the planet, for each postulated planet trajectory. Classical approaches are median-based \cite{marois2006} and likelihood based \cite{mugnier2009,ruffio2017}.

The final third step is to compute the detection map for the image, which assigns a value to every pixel indicating how much we believe a planet is located in the pixel. From the detection map, pixels can then be predicted positive or negative according to whether the value is above or below a detection threshold. A common detection map based on the flux map is the signal-to-noise ratio (denoted by SNR or S/N), which uses a two-sample $t$-test to compare the flux of the considered resolution element to the noise estimated over the remaining resolution elements over an annulus~\cite{mawet2014}. Other methods, such as the STIM detection map \cite{pairet2019} or the regime-switching model (RSM) \cite{dahlqvist2020}, employ more complex models that take into account the possible planet trajectories or the behaviour of quasi-static speckles.

Considering AnnPCA for the first step, median or maximum likelihood for the second step, and SNR for the third step yields two baseline methods that we term \textit{median-SNR} and \textit{L-SNR}, respectively. Our L-SNR is similar to the method proposed in~\cite{mugnier2009,ruffio2017} in that they are based on likelihood and SNR. However, it differs from the work of \cite{mugnier2009,ruffio2017} by assuming a Laplacian instead of Gaussian distributed* noise and by performing the background subtraction by AnnPCA instead of the difference of frames method \cite{mugnier2009} and the Karhunen-Loève image processing method \cite{soummer2012}.

\begin{figure}[t]
    \centering
    \begin{subfigure}{0.45\textwidth}
        \includegraphics[width=\textwidth]{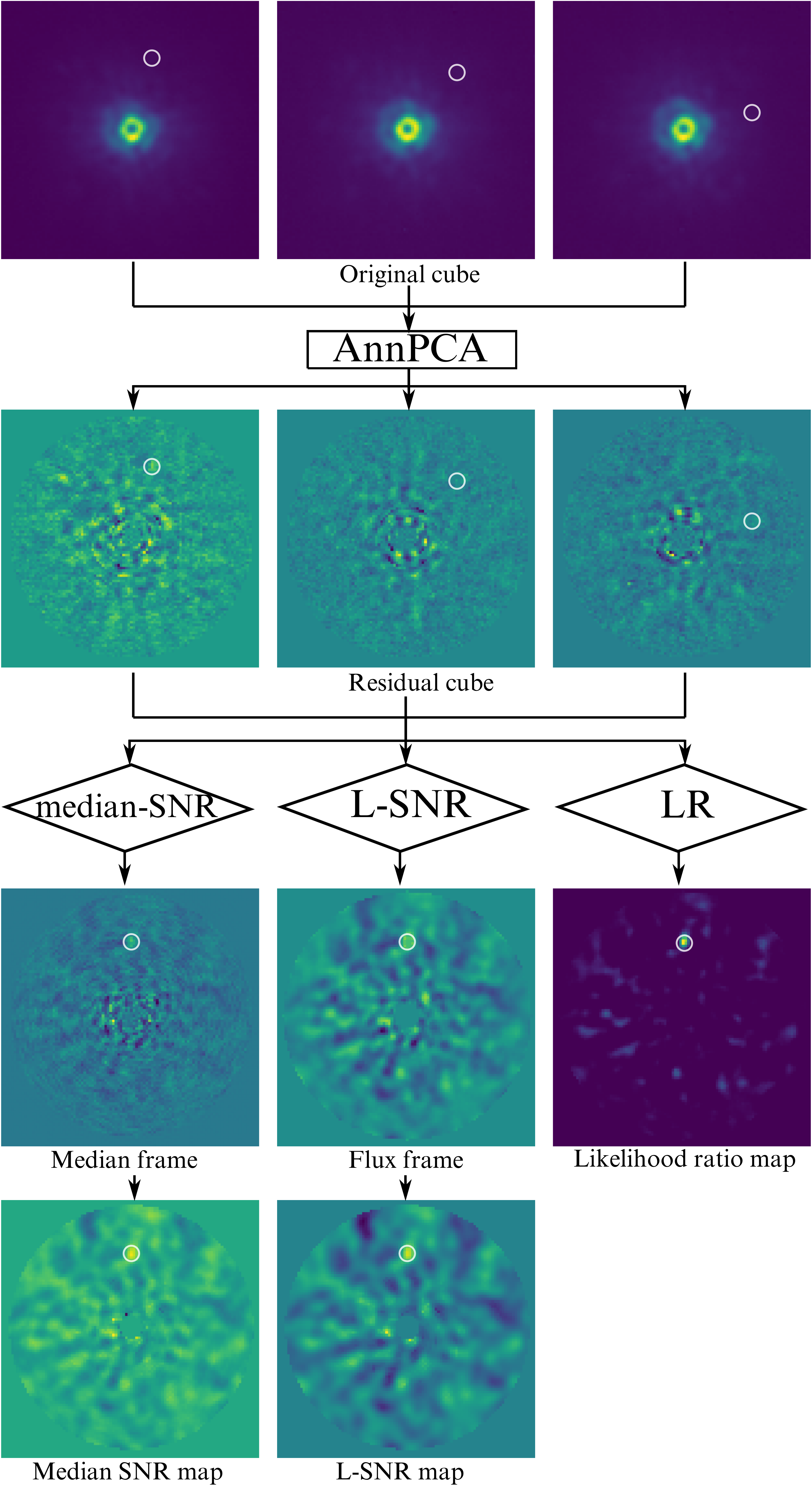}
    \end{subfigure}
    \caption{Pipeline of the algorithms shown on the $\beta$-pictoris datacube with injected planet located at the white circle. For AnnPCA we choose rank $20$.}
\label{fig:pipeline}
\end{figure}

In this paper, we propose to produce a detection map directly from the residual cube by means of a likelihood-ratio approach, which assumes that the noise after the background subtraction step obeys a Laplacian distribution; see Fig. \ref{fig:pipeline} for an illustration of the three considered pipelines. We report results where, in comparison with median-SNR and L-SNR, the likelihood-ratio (LR) approach yields a considerably higher true-positive-rate/false-positive-rate ratio in the high threshold regime.

The paper is structured as follows. Our proposed algorithm for direct exoplanet detection is described in Section \ref{sec:algorithm}. The experimental results are presented in Section \ref{sec:results} to validate the performance of the proposed detection map. Final remarks are given in Section \ref{sec:conclusions}.

\section{Detection based on likelihood ratio map}
\label{sec:algorithm}
Let $R\in\mathbb{R}^{T\times N \times N}$ be a residual cube that is the result of background subtraction, e.g.~using AnnPCA from a sequence of $T$ images of size $N\times N$, and should contain only the planet rotating on an unknown trajectory $g\subset[T] \!\times\! [N]^2$ with some small residual noise term $E$.

We develop a statistical model that postulates that a planet could be located at any pixel of the first frame, which also defines its trajectory $g$, and computes the optimal flux $a$ such that the log-likelihood of observing the noise term $E$ is maximised.

\begin{figure}[t]
    \centering
    \begin{subfigure}{0.35\textwidth}
        \includegraphics[width=\textwidth]{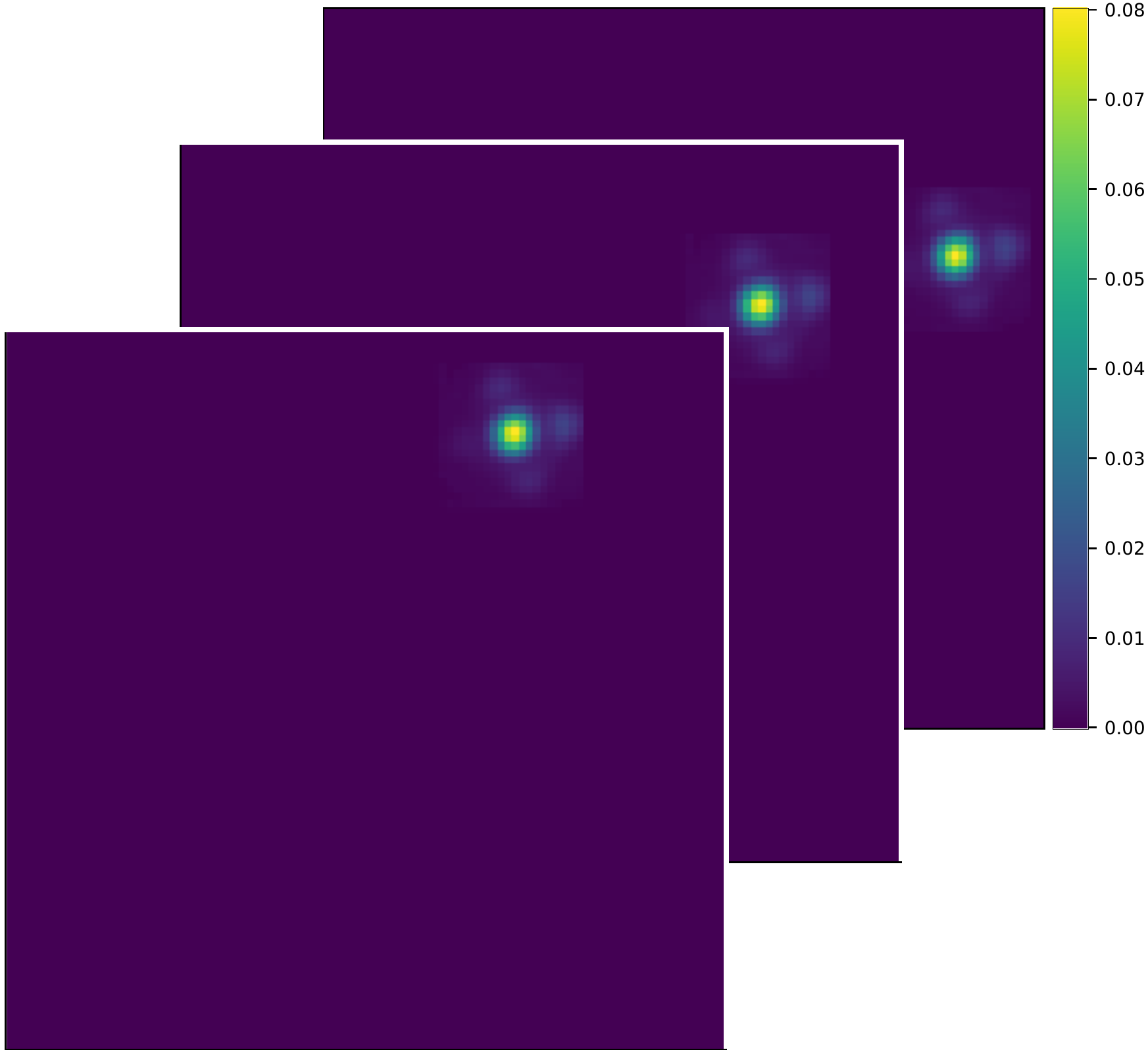}
    \end{subfigure}
    \caption{The cube of the planet signature constructed by rotating position of the PSF function along the trajectory $g$.}
    \label{fig:pg}
\end{figure}

\begin{figure}[t]
    \centering
    \begin{subfigure}{0.35\textwidth}
        \includegraphics[width=\textwidth]{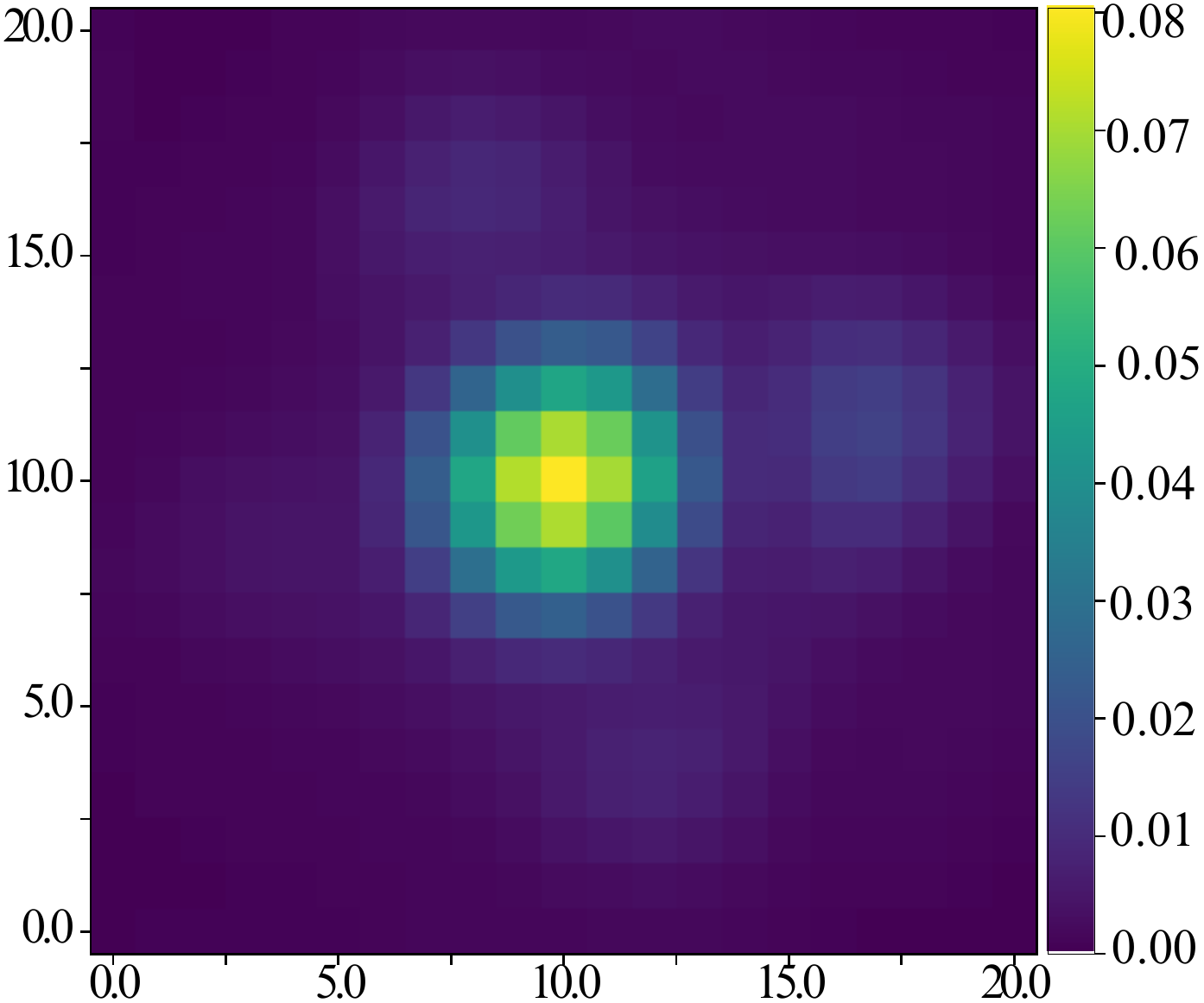}
    \end{subfigure}
    \caption{Normalized point spread function (PSF) of $\beta$ Pictoris data set defined by the optical instrument that was used to acquire the dataset.}
\label{fig:psf}
\end{figure}

The method can be summarised as follows. Assuming there is a planet along the trajectory $g$, the residual cube $R$ is modelled as:
\begin{equation}
\label{eq:rescube_model}
	R = a P_g + E,
\end{equation}
where $a \geq 0$ is the flux, $P_g$ is a cube constructed by rotating the planet signature along $g$ illustrated in Fig.~\ref{fig:pg}, and $E$ is the residual noise. To construct $P_g$ we place a copy of the normalized reference point spread function (PSF) defined by the optical instrument, depicted in Fig.~\ref{fig:psf}, at coordinates $g_t$ on each frame $t$ of a cube with zero entries.

Mugnier et al.~\cite{mugnier2009} propose to model the noise term $E$ as white and Gaussian. It is then possible to estimate the value of $a_g$ by maximizing the following log-likelihood
\begin{equation} \label{eq:mc_gaussian_loglike}
	\log \mathcal{L}^\mathrm{Gauss}_g(a|R) \propto
	-\frac{1}{2} \!\!\!\sum_{(t, r) \in \Omega_g}
	\frac{|R(t,r) - a P_g(t,r)|^2}{\sigma^2_{R(r)}},
\end{equation}
where $\sigma_{R}^2$ is the empirical variance of the residual frames computed along the time dimension and $\Omega_g$ is the set of indices $(t, r)$ of pixels whose distance from the trajectory $g$ is smaller than half the diffraction limit 
\begin{equation} 
\label{eq:listind}
	\Omega_g = \left\{
		(t, r) \in [T] \!\times\! [N]^2
		\,\bigg\rvert\,
		\|r - g_t\|_2 < \frac{1}{2} \frac{\lambda}{D}
	\right\}.
\end{equation}

Maximizing~\eqref{eq:mc_gaussian_loglike} translates to a simple linear least-squares problem \cite{mugnier2009}.

\begin{figure*}[t!]
    \centering
    \begin{subfigure}{0.95\textwidth}
        \includegraphics[width=\textwidth]{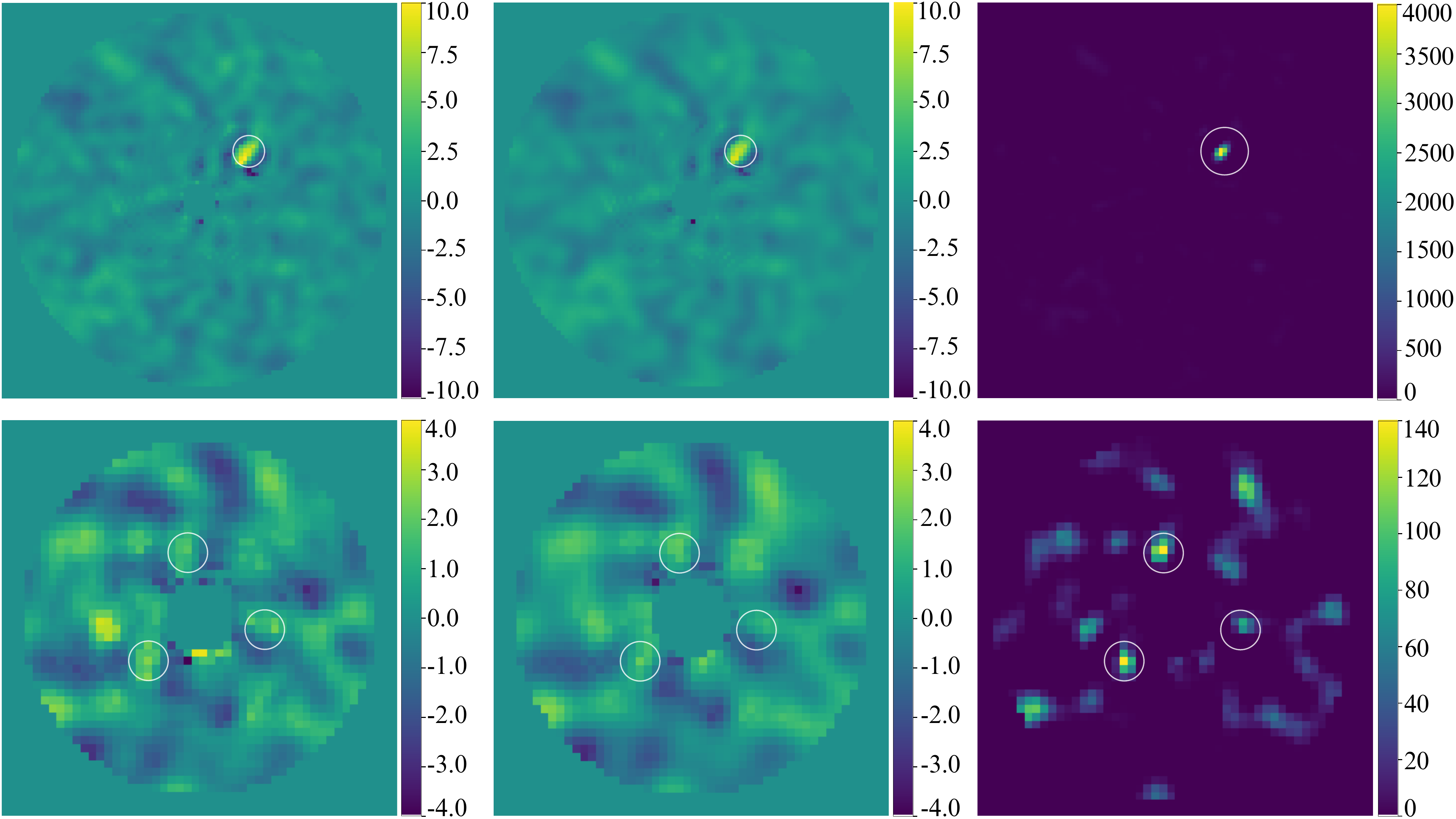}
    \end{subfigure}
\caption{From left to right, median-SNR map, L-SNR map, and LR map, respectively. Firstly, they are tested using the real planet (top line), then using three fake planets injected to the separation 1--2 $\lambda/D$ with $1\sigma_{ann}$ (bottom line). They are obtained with AnnPCA using 20 PCs on $\beta$ Pictoris data set. Note that the simultaneous injection of three fake planets is only for illustration; the ROC curves of Fig.~\ref{fig:roccurves} are obtained with single planet injections. We relied on the VIP package~\cite{gonzalezvip2017} for step 1 and step 3 of the pipeline and for displaying the SNR maps.}\label{fig:detectionmaps}
\end{figure*}

However, it has been observed that the tail of the distribution of the noise term $E$ decays exponentially and is more accurately described by the Laplacian distribution \cite{pairet2019}. We replace the Gaussian assumption in \eqref{eq:mc_gaussian_loglike} with a Laplacian
\begin{equation} \label{eq:mc_lap_loglike}
	\log \mathcal{L}_g(a|R) \propto 
	-\!\!\!\sum_{(t, r) \in {\Omega}_g}
	\frac{|{R(t, r) - a P_g(t, r)}|}{\sigma_{R}(r)},
\end{equation}
which leads to the following optimisation problem that estimates the planet flux
\begin{align} 
\label{eq:mc_lap_loglike_optim}
	\hat{a}_g
	&= \mathrm{argmax}_a \log \mathcal{L}_g(a|R) \\
	&= \mathrm{argmin}_a \sum_{(t, r) \in {\Omega}_g}\!\!\!\!
	\frac{|{R(t, r) - a P_g(t, r)}|}{\sigma_{R}(r)}.
\end{align}

Solving \eqref{eq:mc_lap_loglike_optim} is an instance of the weighted least absolute deviation (LAD) problem, which, unlike least squares, does not have a closed form solution. In general, $\ell_1$ minimization can be solved by a number of efficient iterative methods, however, in our specific case, it is possible to compute the solution even more efficiently. Since the objective function is a convex piecewise linear function $\mathbb{R} \rightarrow \mathbb{R}$ with intervals between $R(t,r)/P_g(t,r)$, $(t, r) \in {\Omega}_g$, its minimum is attained at one of the $(TN^2)$ points that can be easily searched exhaustively.

We propose a new detection map, termed \textit{LR map}, which compares how well two hypotheses, the null ($H_0$) and research ($H_1$) hypotheses, fit the data \cite{casella2001}. In our case, for each trajectory, $H_0$ corresponds to the absence of a companion, and $H_1$ to the presence of one. We define the likelihood ratio $\Lambda(R)$ as the ratio of the maximum likelihood to the likelihood of the null hypothesis and compute its logarithm
\begin{align}
	\log \Lambda_g(R) &= \log \left(\frac{
		 \mathcal{L}_g(\hat{a}_g | R)
	}{
	 \mathcal{L}_g(0 | R)
	}\right) \\
	&= -\!\!\!\!\!\sum_{(t, r) \in {\Omega}_g}\!\!\!\!
		\frac{
			|R(t, r) - \hat{a}_g P_g(t, r)|
			- |{R(t, r)}|
		}{
			\sigma_{R}(r)
		}.
\end{align}
The LR map is then the map which assigns the value $\log \Lambda_g(R)$ to the starting pixel of every trajectory $g$.

\section{Numerical experiments}
\label{sec:results}

We compare the proposed LR detection map with the baseline median-SNR and L-SNR detection maps in terms of visual quality and Receiver Operating Characteristic (ROC) curves. We provide Python code based on the VIP HCI package \cite{gonzalezvip2017} of our experiments publicly\footnote{https://github.com/hazandaglayan/likelihoodratiomap}.

\sloppy
We test the efficacy of our method on the commonly used ADI cube VLT/NACO $\beta$-Pictoris in the infra-red L' band (3.8{$\mu$m}), which  has 612 frames covering 83$^{\circ}$  of parallactic angles and $\lambda/D \approx 4.6\,\text{pixel}$ \cite{absil2013}. In order to reduce the computation time, we consider only every third frame of the data set and crop it to 60 by 60 pixels, resulting into a cube of size $204\times60\times 60$. In AnnPCA, we choose 20 as the number of components, which has been observed to perform well \cite{dahlqvist2020}. To generate synthetic groundtruth examples, we inject the planet-free data cube (i.e., the data cube where the known planets have been removed) with fake planets using the VIP HCI package \cite{gonzalezvip2017}.

Fig.~\ref{fig:detectionmaps} shows the visual quality of detection maps for two cases, the top row is the real data set with an actual planet, and the bottom row is the synthetically produced data set which we acquire by injecting three planet companions into an empty $\beta$-Pic cube.  

We propose a novel deterministic approach to compute the ROC curves from synthetic planet injections. Our approach has the advantage of producing a diagonal ROC curve as expected when the flux of the fake planets is reduced to zero.

\begin{figure*}[t!]
    \centering
    \includegraphics[width=.75\textwidth]{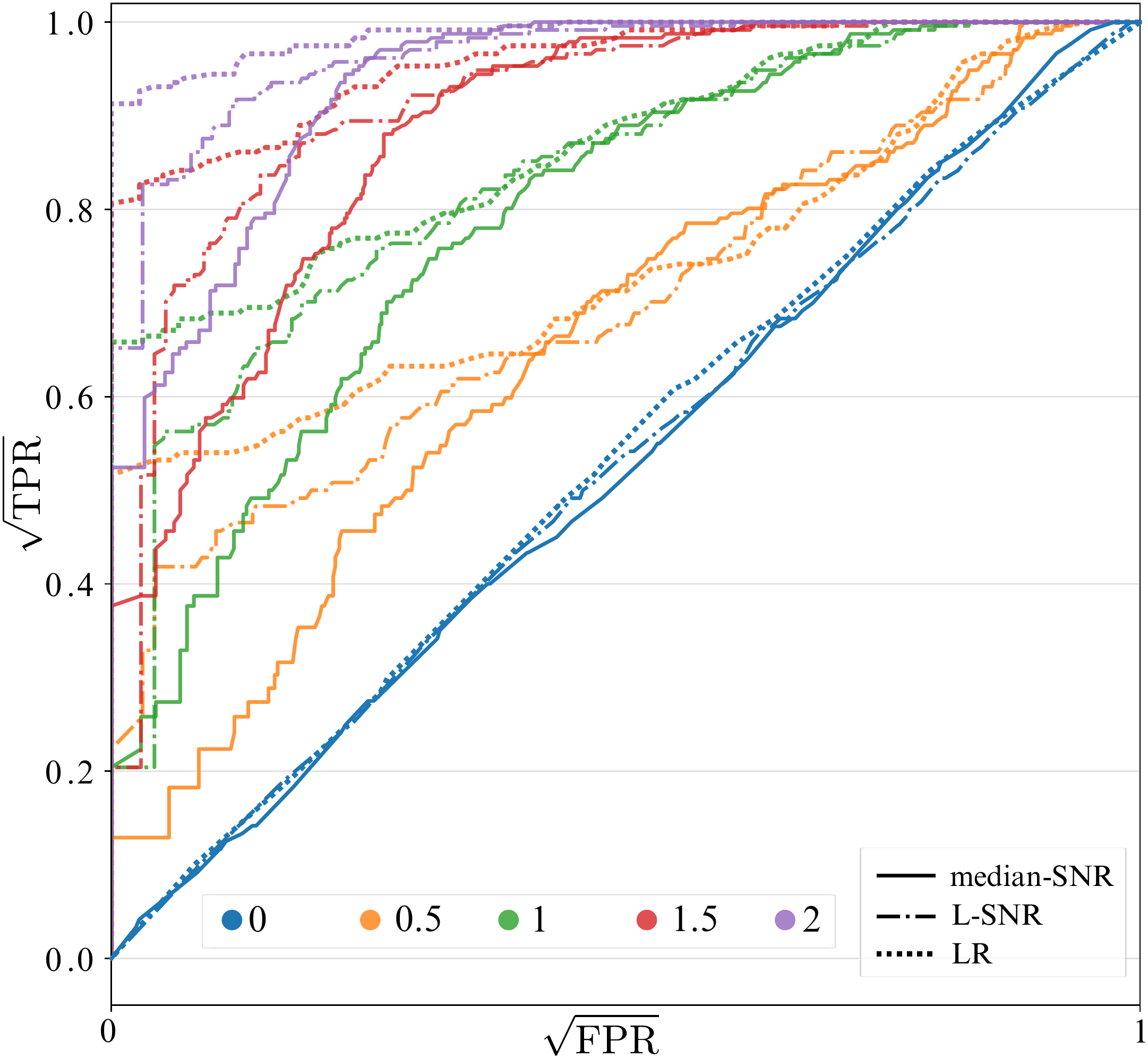}
    \caption{ROC curves for $\beta$ Pictoris data set. The planets are injected in 1--2 $\lambda/D$ separation with the fluxes indicated in the legend multiplied by the standard deviation of the annulus. In order to better observe the low FPR regime, we scale the axes using the square root.}\label{fig:roccurves}
\end{figure*}

Our procedure involves data cubes in which we choose one ``actual positive'' location where we inject a fake planet with prescribed flux, and several ``actual negative'' locations where we inject nothing. The detection algorithm receives as input all those locations, and it applies the exact same detection procedure to all the locations. Specifically, when at least one actual positive resolution element (of size $\lambda/D$) contains an above-threshold pixel of the detection map (median-SNR, L-SNR or LR map), we count one true positive (TP); when at least one actual negative resolution element contains an above-threshold pixel, we count one false positive (FP). This finally yields the true positive rate (TPR) and the false positive rate (FPR) that constitute the vertical and horizontal axes of the ROC curve.

Fig.~\ref{fig:roccurves} depicts the performance of the methods in terms of deterministic ROC curves computed for values of flux relative to the standard deviation of the annulus $c \cdot \sigma_{ann}$, where $c = 0, 0.5, 1, 1.5, 2$. We observe the performance of the LR map to be above the median-SNR and the L-SNR curves in the low FPR region.  Specifically, as the threshold is lowered, more actual positives are detected with LR than SNR before the first actual negatives are detected. 

Table \ref{tab1:maxy} shows the highest TPR at a zero FPR for injected planets with different flux values, which reflects the fraction of detected actual planets  before the first detection of an actual negative location. False positives are troublesome in high-contrast image processing because they require follow-up observations to be disproved, which are very costly in terms of telescope time. Therefore, compared to the more standard area under curve (AUC) metric for ROC curves, we find the highest TPR at zero FPR to be more aligned with the goal of detecting exoplanets. The values in the table show that LR outperforms the other methods.

\begin{table}[htbp]
    \caption{Highest TPR at zero FPR.}
    \begin{center}
        \begin{tabular}{|c|c|c|c|}
            \hline
            \textbf{}&\multicolumn{3}{|c|}{\textbf{Detection Maps}} \\
            \cline{2-4} 
            \textbf{$c$} & \textbf{\textit{median-SNR}}& \textbf{\textit{L-SNR}}& \textbf{\textit{LR}} \\
            \hline
            0& 0& 0& 0 \\
            0.5& 0.017 & 0.050 & \textbf{0.267} \\
            1 & 0.042 & 0.042 &	\textbf{0.433} \\
            1.5 & 0.142 &	0.042 & \textbf{0.633} \\
            2 & 0.275 & 0.425 & \textbf{0.833} \\
            \hline
        \end{tabular}
        \label{tab1:maxy}
    \end{center}
\end{table}

\section{Conclusions}
\label{sec:conclusions}
We have presented a new detection map based on a likelihood ratio for direct imaging of exoplanets which assumes an underlying Laplacian distribution on the noise. We have compared the algorithm using a new ROC curve methodology, which maintains an important property, that for a zero flux injections it returns a diagonal line. Numerical experiments on $\beta$-Pictoris dataset hint at the improvement in terms of detection of exoplanets after applying a widely used Annular PCA, especially in the low FPR regime compared to the other tested detection maps.

\vspace{12pt}

\end{document}